\documentclass[aps,prd,floats,twocolumn,showpacs,floatfix,nofootinbib]{revtex4}
\usepackage{epsfig}

\begin{document}
\draft
\title{An analysis of a QND speed-meter interferometer}

\author{Patricia Purdue}

\address{
Theoretical Astrophysics, California Institute of Technology,
Pasadena, CA 91125
}

\date{2 May 2002}
\begin{abstract}

In the quest to develop viable designs for third-generation
optical interferometric gravitational-wave detectors (e.g. LIGO-III
and EURO), one strategy is to monitor the relative 
momentum or speed of the test-mass mirrors, rather than monitoring
their relative position.  This paper describes and
analyzes the most straightforward design for a {\it speed meter
interferometer} that accomplishes this --- a design (due to 
Braginsky, Gorodetsky, Khalili and Thorne) that is analogous
to a microwave-cavity speed meter conceived by Braginsky
and Khalili.  A mathematical mapping between the microwave
speed meter and the optical interferometric speed meter is developed
and is used to show (in accord with the speed being a Quantum 
Nondemolition [QND] observable) that
{\it in principle} the interferometric speed meter can beat the 
gravitational-wave standard quantum limit (SQL) by 
an arbitrarily large amount, over an arbitrarily wide range
of frequencies, and can do so without the use of squeezed vacuum or 
any auxiliary filter cavities at the interferometer's input or output.  
However, {\it in practice}, to reach or
beat the SQL, this specific speed meter requires exorbitantly
high input light power.  The physical reason for
this is explored, along with other issues such as constraints on
performance due to optical dissipation.  This analysis forms a foundation
for ongoing attempts to develop a more practical variant of an
interferometric speed meter and to combine the speed meter concept
with other ideas to yield a promising LIGO-III/EURO interferometer
design that entails low laser power.

\end{abstract}
\pacs{PACS numbers: 04.80.Nn, 95.55.Ym, 42.50.Dv, 03.67.-a}

\maketitle


\section{Introduction and Summary}
\label{sec:Introduction}

The first generation of kilometer-scale interferometric
gravitational-wave detectors (LIGO-I \cite{ligoIa,ligoIb}, VIRGO \cite{virgo},
and GEO600 \cite{geo})
will begin operation in 2002, at sensitivities where it is
plausible but not highly probable that gravitational waves
can be detected.  Vigorous research and development is now underway for 
second-generation detectors (LIGO-II \cite{ligoII} and its European and Japanese
partners \cite{euro,tamalcgt}) that are planned to begin operation in 
$\sim 2008$ at a sensitivity where a rich variety of 
gravitational-wave sources should lie.

This second-generation sensitivity will be near or modestly 
better than the {\it standard quantum limit} (SQL), a limit that
constrains interferometers \cite{CavesSQLforIFOs} such as 
LIGO-I which have conventional 
optical topology, but does not constrain more 
sophisticated ``quantum nondemolition''
(QND) interferometers \cite{Unruhsqueezedvacuum,JR}.

Conceptual-design R\&D is now underway, at a modest level, for 
third-generation gravitational-wave interferometers that (it is hoped)
will beat the SQL by a factor $\sim$5 or more over a frequency
band somewhat greater than the typical frequency of operation.
This third-generation R\&D has entailed, thus far, conceiving
and exploring theoretically a number of ideas that might prove useful
in a final design.  Examples include (i) injecting squeezed vacuum into
an interferometer's dark port 
\cite{Unruhsqueezedvacuum,JR,Cavessqueezedvacuum}, (ii) performing homodyne
detection on the output light with frequency-dependent homodyne angle 
(achieved using large Fabry-Perot filter cavities) \cite{freqdephomodyne,Kimble}, 
(iii) using light pressure to transfer the gravity-wave signal onto a small
test mass that moves relative to local inertial frames and then reading
out that motion using local QND techniques (the {\it Optical Bar}) 
\cite{opticalbar}, (iv) a variant of this involving {\it Symphotonic States} 
\cite{symphotonicstates}, (v) producing {\it Optical-Spring} behavior by means 
of a signal-recycling mirror \cite{opticalspring}, and (vi) other more 
general means of changing the dynamics of the test-mass mirrors 
\cite{otherligoIIIa,otherligoIIIb}.  

The purpose of this paper is to carry
out a first detailed analysis of another idea that may prove helpful
in third-generation interferometers: operating each interferometer
as a {\it speed meter}, so instead of monitoring the relative position
of its test-mass mirrors, it measures their relative speed (or, more
precisely, some combination of their speed and higher-order time derivatives
of relative position). 

The motivation for measuring speed rather than position, stated in somewhat
heuristic terms, is as follows:  
If a single measurement of the relative position of the test masses
is made, then according to the uncertainty principle, there
will be a corresponding random ``kick" to the relative momentum.  This
kick will
affect the future positions of the test masses.  If another
position measurement is made at a later time, its accuracy
will be limited because of the earlier momentum kick.  The
best one can do is balance the uncertainties of the two measurements;
this optimal uncertainty corresponds to the SQL.

If, on the other hand, the velocity (which is directly proportional to the
momentum) is measured directly, this velocity measurement
will randomly kick the relative position.
That position kick is irrelevant {\it if}
the velocity is being measured without collecting position information,
as in a speed meter.  Another way to say this is to note that the
velocity (or momentum) is a constant of the free motion of the test mass.
Consequently, the velocity commutes with itself at different times and
is therefore a quantum non-demolition (QND) observable 
\cite{sqlandqndobservables}.
The result is that speed meters are not constrained by the
SQL.  

The original idea for a speed meter that measures the velocity
of a single test mass was conceived, in a primitive form, by
Braginsky and Khalili \cite{firstspeed}.  Braginsky, Gorodetsky, 
Khalili, and Thorne \cite{Brag} (henceforth called BGKT)
devised a refined and marginally practical form based on two coupled
microwave cavities.  In their appendix, BGKT also sketched a design for
an optical-interferometer speed meter gravity-wave detector that, they
speculated, will be able to beat the gravity-wave SQL in essentially
the same manner as the microwave speed meter beats the free-mass SQL.

This paper presents a detailed analysis of the BGKT optical-interferometer
speed meter, with the objective of determining whether it actually does 
measure relative velocity without collecting position information and whether 
it actually can beat the SQL.  As we shall see, the answers are both ``yes."
Moreover, it will be shown that there is a mathematical mapping between
the analysis of the microwave-cavity speed meter, which measures the 
velocity of a single mass, and that of the optical-interferometer 
speed meter, which measures the relative speeds of widely separated 
test masses.   Another objective of this paper is to explore the 
features of this optical-interferometer speed meter that will be important
in attempts to design practical third-generation interferometers.

\begin{figure}
\epsfig{file=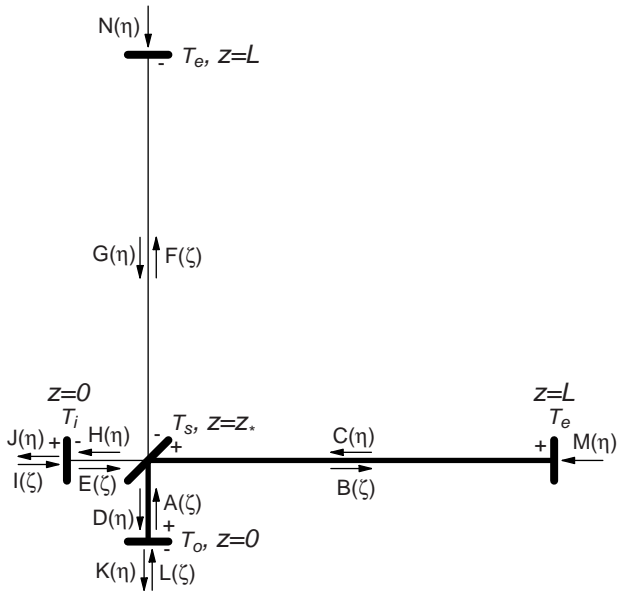}
\caption{Design for QND speed meter interferometer.  The main laser input port is
the lower left mirror [denoted by $I(\zeta)$, where $\zeta=t-z/c$].  The signal 
is extracted at the bottom mirror [denoted $K(\eta)$, where 
$\eta=t+z/c$].  The ``$+$" and ``$-$" signs near the mirrors indicate the sign of
the reflectivities in the junction conditions for each location.}
\label{fig:schematic}
\end{figure}

The basic design of the speed meter to be analyzed here is shown
in Figure \ref{fig:schematic}. 
It consists
of two nearly identical optical cavities of length $L=4{\rm km}$, which 
are weakly coupled by a mirror of power transmissivity $T_{\rm s}$.
In the absence of a driving force, laser light can ``slosh" back
and forth between these two cavities with frequency 
\begin{equation}
\Omega=c\sqrt{T_{\rm s}}/L \;,
\label{slosh}
\end{equation} 
where $c$ is the speed of light.
The addition of a driving laser [denoted $I(\zeta)$ in
Fig.\ \ref{fig:schematic}]
into one cavity will cause the other cavity to become 
excited.  It is from this excited cavity that we will 
extract our signal [denoted $K(\eta)$] at a rate 
\begin{equation}
\delta = cT_{\rm o}/4L \;,
\label{extract}
\end{equation}
where $T_{\rm o}$ is the transmissivity
of the extraction mirror.  Since we cannot make $T_{\rm o}$
infinitely small
(or equivalently, the extraction time infinite), a small amount of
residual light will build up in the unexcited cavity.  To counteract
this, we also input a small
amount of laser light [denoted $L(\zeta)$] through the
output port in order to cancel out any such residual light.
This is desirable because one 
cavity must be empty to achieve pure speed meter behavior.\footnote{In
general, one could allow some amount of light to build up in the ``empty"
cavity, and thereby (as we shall see in Sec.\ \ref{sec:losslessdisc}),
make it easier to inject light into the interferometer.  Then, the 
ratio of the levels of excitation of the two cavities
would become a tool for optimizing the design, balancing reduced input 
power against performance.}

To understand how this system produces a velocity signal, consider
the effect of moving the end mirror in the excited cavity [the cavity
labeled $C(\eta)$ and $B(\zeta)$ in Fig. \ref{fig:schematic}].  That 
mirror motion will put a phase shift on the light in that cavity.  
If the input laser is driving the cavity's cosine quadrature,
then the phase shift caused by the mirror motion will act as a driving
force for the sine quadrature.  This light will then slosh into the 
empty cavity and back.  When it returns, it will be $180^\circ$ out of phase
compared to its initial phase shift.  The resulting cancellation will cause
the net signal in the sine quadrature of the excited cavity to
be proportional to the difference in test-mass position between 
the start and finish of the sloshing cycle.  In other words, 
the net signal is proportional to the velocity of the test mass, 
assuming that the frequencies $\omega$ of the test mass' motion 
are $\omega \ll \Omega$.

As it turns out, however, the optimal regime of operation for the 
speed meter is $\omega \sim \Omega$.  Consequently, the output signal
contains a sum over odd time derivatives of position
[see Eq.\ (\ref{kPhiOut}) and the discussion following it].  
Therefore, the speed meter monitors not just the
relative speed of the test masses, but a mixture of
all odd time derivatives of the position.

As we will show, this speed meter design, {\it in principle}, 
is capable of beating the
SQL by an arbitrary amount and over a wide range of frequencies.  
However, {\it in practice}, optical losses will limit the amount by 
which the SQL can be beaten, and to beat the SQL at all requires an 
uncomfortably high circulating power.  (This is actually
a common feature of designs that beat the SQL \cite{highpower}.)  
More seriously, this design requires an impossibly high input power because
the photons are not getting ``sucked" into the interferometer efficiently,
as they are in conventional designs; this will be discussed in more detail
in Sec.\ \ref{sec:losslessdisc}.

In view of this impracticality, one might wonder why a detailed analysis 
of this speed meter should be published.  The answer is that this analysis 
teaches us a variety of things about optical-interferometer speed meters --- 
things that are likely to be of value in the search for practical QND 
interferometers and in their optimization.  Indeed, the author and 
Yanbei Chen are now exploring more sophisticated and promising 
speed-meter designs that rely, for motivation and insights, on the 
things learned in the analysis presented here.

This paper is organized as follows:
The mathematical description of the interferometer is given in
Sec.\ II.  Sec.\ \ref{sec:losslessmath} gives the analysis of the 
lossless limit and the 
mapping to the microwave-resonator speed meter,  Sec.\ \ref{sec:losslessnum} 
presents the
numerical analysis, and as mentioned above, Sec.\ \ref{sec:losslessdisc}
describes various 
problems or issues with the speed meter.  In Sec.\ \ref{sec:lossy}, we
give the results and a description of the speed meter's
performance if losses are included.  The discussion there will address
the role of optical dissipation in limiting the amount by which the 
SQL can be beaten.  Finally, Sec.\ V summarizes the
results of this analysis and its relevance for future research.

\section{Mathematical Description of the Interferometer}
\label{sec:math}

The design of the speed meter is shown in Figure \ref{fig:schematic}.
In this section, we will set up the equations describing the
interferometer with lossy mirrors. The method of analysis is
based on the formalism developed by Caves and Schumaker \cite{cs}
and used by Kimble, Matsko, Thorne, and Vyatchanin (KLMTV)
\cite{Kimble} to examine more conventional interferometer designs.

We express the electric field propagating in each direction down each
segment of the interferometer in the form
\begin{equation}
E_{\rm field}(\zeta) = \sqrt{\frac{4 \pi \hbar \omega_0}{{\cal S} c}} A(\zeta) \;,
\label{field}
\end{equation}
where $A(\zeta)$ is the amplitude, 
$\zeta = t - z/c$ (see Fig.\ \ref{fig:schematic}), $\omega_0$ is the carrier
frequency, $\hbar$ is the reduced Planck's constant, 
and ${\cal S}$ is the effective cross-sectional area of the light
beam; see Eq. (8) of KLMTV.  We decompose the amplitude into cosine and
sine quadratures,
\begin{equation}
A(\zeta) = {\cal A}_1(\zeta) \cos \omega_0 \zeta 
+ {\cal A}_2(\zeta) \sin \omega_0 \zeta \;.
\label{amplitude}
\end{equation}
Note that the subscript 1 always refers to the cosine 
quadrature, and 2 to sine.  Also note that we have designated 
$z=0$ at both the input and output mirrors, $z=z_*$ at the 
sloshing mirror, and $z=L$ at the end mirrors; see Fig.\ \ref{fig:schematic}.
We choose the cavity lengths $L$ to be exact half multiples of the
carrier wavelength so $e^{i 2 \omega_0 L/c} =1$.

As mentioned above, the power transmissivity for the sloshing 
mirror is $T_{\rm s}$ and for the output mirror is $T_{\rm o}$.  In
addition, $T_{\rm i}$ will denote the transmissivity
for the laser-input mirror and $T_{\rm e}$ for the end
mirrors; again, see Fig.\ \ref{fig:schematic}.  Each of these has 
a complementary reflectivity 
such that each mirror satisfies the equation $T+R=1$.  
If we now let $\zeta = t - z/c$, $\eta = t + z/c$, and $j=1,2$, 
then the junction conditions at the mirrors are given by:

\begin{mathletters}
\label{eqs:junction}
\begin{eqnarray}
{\cal A}_j(\zeta) &=& \sqrt{T_{\rm o}}{\cal L}_j(\zeta) + \sqrt{R_{\rm o}}
	{\cal D}_j(\eta) \;,
\label{eq:Aeqn}\\
{\cal B}_j(\zeta) &=& \sqrt{T_{\rm s}}{\cal E}_j(\zeta) + \sqrt{R_{\rm s}}
	{\cal A}_j(\zeta) \;,\\
{\cal C}_j(\eta) &=& \sqrt{T_{\rm e}} {\cal M}_j(\eta)
			+ \sqrt{R_{\rm e}} {\cal B}_j(\zeta) \;,
				\label{eq:Ceqn}\\
{\cal D}_j(\eta) &=& \sqrt{T_{\rm s}}{\cal G}_j(\eta) +
	\sqrt{R_{\rm s}}{\cal C}_j(\eta) \;, \\
{\cal E}_j(\zeta) &=& \sqrt{T_{\rm i}}{\cal I}_j(\zeta)
			- \sqrt{R_{\rm i}}{\cal H}_j(\eta) \;, \\
{\cal F}_j(\zeta) &=& \sqrt{T_{\rm s}}{\cal A}_j(\zeta) - \sqrt{R_{\rm s}}
	{\cal E}_j(\zeta) \;, \\
{\cal G}_j(\eta) &=& \sqrt{T_{\rm e}}{\cal N}_j(\eta)
			- \sqrt{R_{\rm e}}{\cal F}_j(\zeta) \;,
				\label{eq:Geqn} \\
{\cal H}_j(\eta) &=& \sqrt{T_{\rm s}}{\cal C}_j(\eta) - \sqrt{R_{\rm s}}
	{\cal G}_j(\eta) \;, \\
{\cal J}_j(\eta) &=& \sqrt{T_{\rm i}}{\cal H}_j(\eta)
			+ \sqrt{R_{\rm i}}{\cal I}_j(\zeta) \;, \\
{\cal K}_j(\eta) &=& \sqrt{T_{\rm o}}{\cal D}_j(\eta) - \sqrt{R_{\rm o}}
{\cal L}_j(\zeta)\;.
\label{eq:Keqn}
\end{eqnarray}
\end{mathletters}

\subsection{Carrier Light}

If we first consider only the carrier in a steady state, 
we can assume that all the mirrors are stationary and that all
of the ${\cal A}_j(\zeta)=A_j$, ${\cal B}_j(\zeta)=B_j$, etc.\ are 
constant.  (We denote the carrier amplitudes by capital Latin letters
with a subscript indicating the quadrature.)  
Then we solve Eqs.\ (\ref{eqs:junction}) simultaneously.
We ignore vacuum fluctuation noise since it is 
unimportant for the carrier light.  In addition, we only drive the 
cosine quadrature, so that 
\begin{equation}
L_2 = I_2 = 0 \;.
\end{equation}
Thus, all of the sine quadrature terms will be zero.  As mentioned 
above, we want to have as little light as possible in the unexcited 
cavity, so we apply the condition 
\begin{equation}
F_1=G_1=0 \;.
\end{equation}
That means the input fed into the output port should be
\begin{equation}
L_1 = \frac{I_1}{4} \sqrt{\frac{T_{\rm o} T_{\rm i}}{T_{\rm s}}}
\label{eq:darkdriving} \;.
\end{equation}
Then, the solution for the carrier is
\begin{mathletters}
\begin{eqnarray}
A_1 = B_1= C_1 &=& D_1 = \frac{I_1}{2} \sqrt{\frac{T_{\rm i}}{T_{\rm s}}} \;,
\label{eq:carrier} \\
E_1 = H_1 &=& \frac{I_1}{2} \sqrt{T_{\rm i}} \;.
\end{eqnarray}
\label{allcarriers}
\end{mathletters}
In deriving Eqs.\ (\ref{allcarriers}), we have assumed the following inequalities among the various mirror transmissivities:
\begin{equation}
T_{\rm o} \gg T_{\rm s} \gg T_{\rm i} \gg T_{\rm e} \;.
\label{transreq}
\end{equation}
The motivations for these assumptions are that: (i) they lead to speed-meter 
behavior; (ii) as with any interferometer, the best performance is achieved by
making the end-mirror transmissivities $T_{\rm e}$ as small as possible; (iii)
good performance requires a light extraction rate comparable to the sloshing
rate, $\delta \sim \Omega$ [cf.\ the first paragraph of 
Sec.\ \ref{sec:losslessnum}], which with
Eqs.\ (\ref{slosh}) and (\ref{extract}) implies $T_{\rm o} \sim \sqrt{T_{\rm s}}$
so $T_{\rm o} \gg T_{\rm s}$; and (iv) if the input transmissivity is larger
than or of the same order as the sloshing frequency, too much light will be 
lost during the sloshing cycle, resulting in incomplete cancellation of the
position information and degraded performance (hence, we assume
$T_{\rm i} \ll T_{\rm s}$).

\subsection{Sideband Light}

Sidebands are put onto the carrier by the mirror motions and by
vacuum fluctuations, as we shall see below.  We express the quadrature
amplitudes for the carrier plus the side bands in the form
\begin{equation}
{\cal A}_j(\zeta) =
A_j + \int_0^\infty \bigl[ \tilde{a}_j(\omega) e^{-i \omega \zeta}
+ \tilde{a}_j^{\dagger}(\omega) e^{i \omega \zeta}\bigr]
\frac{d\omega}{2\pi} \;,
\label{sidebandamp}
\end{equation}
where $\tilde{a}_j(\omega)$ is the field amplitude for the sideband
at frequency $\omega$ in the $j$ quadrature; cf.\ Eqs.\ (6)--(8) of KLMTV,
where commutation relations and the connection to creation and annihilation
operators are discussed.
Then most of the junction conditions can easily be
broken down into separate expressions for the constant and
sideband terms; for example,
\begin{mathletters}
\label{eqs:Aeq}
\begin{eqnarray}
A_j &=& \sqrt{T_{\rm o}} L_j + \sqrt{R_{\rm o}} D_j \;, \\
\tilde{a}_j &=& \sqrt{T_{\rm o}} \tilde{\ell}_j
	+ \sqrt{R_{\rm o}} \tilde{d}_j \;.
\end{eqnarray}
\end{mathletters}
The exceptions are Eqs.\ (\ref{eq:Ceqn}) and (\ref{eq:Geqn}) because
the two end mirrors will change the phase of the sidebands on each
bounce.  Equation (\ref{eq:Geqn}) becomes
\begin{mathletters}
\label{eqs:Geq}
\begin{eqnarray}
G_j &=& - \sqrt{R_{\rm e}} F_j + \sqrt{T_{\rm e}} N_j \;, \\
\tilde{g}_j &=& - \sqrt{R_{\rm e}} \tilde{f}_j e^{i\beta}
+ \sqrt{T_{\rm e}} \tilde{n}_j \;,
\end{eqnarray}
\end{mathletters}
where $\beta = 2\omega L/c$ is the phase shift for the sidebands.
At this point, we also want to allow mirror motion in order to
detect gravitational waves, so we assume that the
end mirror of the excited cavity is free to move.  As a result, the
junction condition there, expressed by Eq.\ (\ref{eq:Ceqn}),
is the most complicated.  It becomes
\begin{mathletters}
\label{eqs:Ceq}
\begin{eqnarray}
C_j &=& \sqrt{R_{\rm e}} B_j + \sqrt{T_{\rm e}} M_j \;, \\
\tilde{c}_1 &=& \sqrt{R_{\rm e}} \tilde{b}_1 e^{i\beta}
		- 2 \sqrt{R_{\rm e}} B_2 \omega_0 \tilde{x}/c
		+ \sqrt{T_{\rm e}} \widetilde{m}_1 \;, \\
\tilde{c}_2 &=& \sqrt{R_{\rm e}} \tilde{b}_2 e^{i\beta}
		+ 2 \sqrt{R_{\rm e}} B_1 \omega_0 \tilde{x}/c
		+ \sqrt{T_{\rm e}} \widetilde{m}_2 \;,
\end{eqnarray}
\end{mathletters}
where $\tilde{x}$ is the Fourier transform of the mirror's displacement.
(We are ignoring the motion of the end mirror of the empty
cavity since that will not have a significant effect.)

All of the junction condition equations [Eqs.\ (\ref{eqs:junction})
expressed in the form of Eqs.\ (\ref{eqs:Aeq}), (\ref{eqs:Geq}),
and (\ref{eqs:Ceq})] can be solved simultaneously to get
expressions for the carrier and sidebands in each segment of
the interferometer.  This yields an output [$K(\eta)$] containing an
$\omega \tilde{x}$ term, which is the Fourier transform of the end-mirror
velocity (relative to the input mirror), aside from a factor of $i$.  Since
there is no factor $\tilde{x}$ without a multiplying factor $\omega$ in the
output, our interferometer is indeed a speed meter, as claimed by BGKT.

One more complication to be addressed is the issue of the back
action force on the mirror produced by the fluctuating radiation
pressure of the laser beam.  The back action is included in
$\tilde{x}$ along with the gravitational wave information, as follows.

From KLMTV, Eq.\ (B18), the back-action force is
\begin{equation}
F_{BA} =  \frac{2 \delta W_{\rm circ}}{c} \;,
\label{eq:baforce}
\end{equation}
where $\delta W_{\rm circ}$ is the fluctuation in the circulating
laser power.  To determine this quantity, consider the expression
for the circulating power [text above Eq.\ (B16) in KLMTV]:
\begin{equation}
W_{\rm circ} = \frac{\overline{E^2_{\rm int}}}{4 \pi} {\cal S} c \;,
\label{eq:powerformula}
\end{equation}
where ${\cal S}$ is the effective cross-sectional area of the beam and
$\overline{E^2_{\rm int}}$ is the time-averaged square of the
internal electric field.  In our case,
\begin{eqnarray}
E_{\rm int} =&& \sqrt{\frac{4\pi \hbar \omega_0}{{\cal S} c}} \times \nonumber \\
            && \quad \Biggl\{ \cos (\omega_0 t)
              \biggl[ B_1 + \int_0^\infty \bigl( \tilde{b}_1 e^{-i \omega t}
           			+ \tilde{b}_1^\dagger e^{i\omega t} \bigr)
              \frac{d \omega}{2 \pi} \biggr] \nonumber \\
           && \quad + \sin (\omega_0 t)
              \biggl[ \int_0^\infty \bigl( \tilde{b}_2 e^{-i \omega t}
                    + \tilde{b}_2^\dagger e^{i\omega t} \bigr)
              \frac{d \omega}{2 \pi} \biggr]
          \Biggr\} \;.
\label{eq:Eint}
\end{eqnarray}
See Eqs. (\ref{field}), (\ref{amplitude}), and (\ref{sidebandamp})
with $\cal A$ replaced by $\cal B$.
Note that the constant term $B_2$ vanishes since we are driving
only the cosine quadrature.  Substituting Eq.\ (\ref{eq:Eint})
into Eq.\ (\ref{eq:powerformula}) will give a steady circulating
power
\begin{equation}
W_{\rm circ} =  \frac{1}{2} \hbar \omega_0 B_1^2
= \hbar \omega_0 I_1^2 \frac{T_{\rm i}}{8 T_{\rm s}}
\label{eq:poweramp}
\end{equation}
and a fluctuating piece
\begin{equation}
\delta W_{\rm circ} (t)
= \hbar \omega_0 B_1 \int_0^\infty \tilde{b}_1 (\omega) e^{-i \omega t}
\frac{d \omega}{2\pi} + {\rm HC} \;,
\label{eq:circpwr}
\end{equation}
where HC denotes the Hermitian conjugate of the previous term.

Now that we have an expression for $\delta W_{\rm circ}$, we return to
the expression for the back-action force (\ref{eq:baforce}).
That force, together with the gravitational waves, produces a
relative acceleration of the cavity's two mirrors (each with mass $m$)
given by
\begin{equation}
\frac{d^2 x(t)}{dt^2} = \frac{1}{2} L \frac{d^2 h(t)}{dt^2}
+ \frac{4 \delta W_{\rm circ}(t)}{mc} \label{eq:displacement}
\end{equation}
where $h(t)$ is the gravitational-wave field [cf.\ Eq.\ (B19) in KLMTV].
Substituting Eq.\ (\ref{eq:circpwr}) into the
above equation and taking the Fourier transform gives
\begin{equation}
\tilde{x} = \frac{1}{2} L\tilde{h} - \frac{4 \hbar \omega_0 B_1 \tilde{b}_1}
{m c \omega^2} \;.
\label{xb}
\end{equation}
Here $B_1$ is given by
Eq.\ (\ref{eq:carrier}), and $\tilde{b}_1$, as obtained by
solving the junction conditions and simplifying with the
conditions on the transmissivities (\ref{transreq}), is given by
\begin{equation}
\tilde{b}_1 = \frac{-i\omega c\sqrt{T_{\rm o}}\tilde{\ell}_1}
	{2L{\cal L}(\omega)} \;,
\label{b1}
\end{equation}
where
\begin{equation}
{\cal L}(\omega) = \Omega^2 - \omega^2 - i \omega \delta \;.
\label{calL}
\end{equation}
(Recall that $\Omega = c\sqrt{T_{\rm s}}/L$ is the sloshing frequency,
$\delta = cT_{\rm o}/4L$ the extraction rate.)

\section{Speed Meter in the Lossless Limit}
\subsection{Mathematical Analysis}
\label{sec:losslessmath}

For simplicity, in this section we will set $T_{\rm e}=0$ (end
mirrors perfectly reflecting), since it is unimportant if
$T_{\rm e}$ is much smaller than the other transmissivities.  We will
also neglect the noise coming in the main laser port ($\tilde{i}_{1,2}$).
This noise will become dominant at sufficiently low frequencies
(below $\sim$10 Hz for the interesting parameter regime), but those
frequencies are not very relevant to LIGO.

As a result of these assumptions, the only noise
that remains is that which comes in through the output port
($\tilde{\ell}_{1,2}$). An interferometer in which this is the case
and in which light absorption and scattering are unimportant ($R+T=1$
for all mirrors, as we have assumed) is said to be ``lossless."
In Sec.\ \ref{sec:lossy}, we shall relax these assumptions; i.e.\ we
shall consider lossy interferometers. As before, we assume
$T_{\rm o} \gg T_{\rm s} \gg T_{\rm i}$.  The interferometer output,
as derived by the analysis of the previous section, is then
\begin{mathletters}
\begin{eqnarray}
\tilde k_1 &=& - {{\cal L}^*(\omega)\over{\cal L}(\omega)}\tilde \ell_1 \;,
\label{k1out} \\
\tilde{k}_2 &=&
   \frac{-i \omega \omega_0 I_1 \sqrt{T_{\rm o} T_{\rm i}}}{2L \sqrt{T_{\rm s}}
    {\cal L}(\omega)} \tilde{x}
   -\frac{{\cal L}^\ast (\omega)}{{\cal L}(\omega)} \tilde{\ell}_2\;,
\label{k2out}
\end{eqnarray}
\label{ksout}
\end{mathletters}
where the asterisk (in ${\cal L}^\ast (\omega)$) denotes the
complex conjugate.  Note that $\tilde{x}$ is given by Eq.\ (\ref{xb})
combined with Eqs.\ (\ref{eq:carrier}) and (\ref{b1}), or equivalently,
by
\begin{equation}
\tilde{x} =  \frac{1}{2} L\tilde{h} + \tilde{x}_{\rm BA} \;,
\label{xBA}
\end{equation}
where
\begin{equation}
\tilde{x}_{\rm BA}
   = \frac{i \hbar \omega_0 I_1 \sqrt{T_{\rm o} T_{\rm i}} \tilde{\ell}_1}
   	{m \omega L \sqrt{T_{\rm s}} {\cal L}(\omega)}
\label{BA}
\end{equation}
is the back-action noise.
It is possible to express Eqs. (\ref{ksout}) in a more concise form,
similar to Eqs.\ (16) in KLMTV:
\begin{mathletters}
\begin{eqnarray}
\tilde{k}_1 &=& \tilde{\ell}_1 e^{2 i \psi} \;, \\
\tilde{k}_2 &=& (\tilde{\ell}_2-\kappa \tilde{\ell}_1) e^{2i \psi}
	+ \sqrt{\kappa} \frac{\tilde{h}}{h_{SQL}^{\rm conv}} e^{i \psi} \;,
\end{eqnarray}
\label{ksoutKimbleform}
\end{mathletters}
where
\begin{mathletters}
\begin{eqnarray}
\tan \psi &=& - \frac{\Omega^2-\omega^2}{\omega \delta}\;, \\
\kappa &=& \frac{\hbar \omega_0^2 I_1^2 T_{\rm o} T_{\rm i}}
	{2mL^2 T_{\rm s} |{\cal L}(\omega)|^2}\;,
\end{eqnarray}
\label{Kimbleparams}
\end{mathletters}
and
\begin{equation}
h_{SQL}^{\rm conv} = \sqrt{\frac{8 \hbar}{m \omega^2 L^2}} \label{sqlconv}\;.
\end{equation}
If, as in KLMTV, we regard Eqs.\ (\ref{ksoutKimbleform}) as input-output
relations for the interferometer, then $\kappa$ is a dimensionless coupling
constant, which couples the gravity wave signal $\tilde{h}$ into the
output $\tilde{k}_2$, $\tilde{h}_{SQL}^{\rm conv}$ is the standard quantum
limit for a conventional interferometer such as LIGO-I, and $\psi$ and $\varphi$
are phases put onto the signal and noise by the interferometer.
Although there is much similarity between the above equations
(\ref{ksoutKimbleform}) and those
of KLMTV, there is not a direct mapping because KLMTV analyzes a
position meter, not a speed meter.

As a tool in optimizing the interferometer's performance, we perform
homodyne detection on the outputs $\tilde{k}_1$ and $\tilde{k}_2$, using
a constant (frequency-independent) homodyne angle $\Phi$.  In other
words, we read out
$\tilde{k}_\Phi = \tilde{k}_1\cos\Phi + \tilde{k}_2\sin\Phi$. If we insert
Eqs.\ (\ref{ksout}) and do some
algebra, we get:
\begin{equation}
\tilde k_\Phi =
   \frac{-i \omega \omega_0 I_1 \sqrt{T_{\rm o} T_{\rm i}}}
   	{2L \sqrt{T_{\rm s}} {\cal L}(\omega)}
\sin\Phi [\tilde x(\omega)+\tilde x_m(\omega)]\;.
\label{kPhiOut}
\end{equation}
Here $\tilde x_m$, the measurement noise (actually shot noise), is given by
\begin{equation}
\tilde{x}_m = {2L\sqrt{T_{\rm s}}{\cal L}^*(\omega)\over
	i\omega\omega_0 I_1 \sqrt{T_{\rm o} T_{\rm i}}}
 [\tilde{\ell}_2 + \tilde{\ell}_1 \cot\Phi ]\;,
\label{eq:xm}
\end{equation}
and  $\tilde x$ is given by Eqs.\ (\ref{xBA}) and (\ref{BA}). Notice that the
first term in Eq. (\ref{kPhiOut}) contains $\tilde{x}$ only in the form
 $\omega \tilde{x}$;
this is the velocity signal [actually,
the sum of the velocity and higher odd time derivatives of position because of
the ${\cal L}(\omega)$ in the denominator].  These
equations, (\ref{kPhiOut}) and (\ref{eq:xm}), are equivalent
to Eqs.\ (29) and (30) of BGKT.
In fact, the analysis of the single-test-mass, microwave speed meter in that
reference (Sec.\ \ref{sec:losslessdisc}) can be translated more or less
directly into the analysis of our
speed-meter interferometer with a suitable change of notation
(see Table \ref{mapping}).\footnote{There is a slight difference
in the way the models in this paper and in BGKT were defined.
One result is that there
are some sign and quadrature differences between them.  For details,
see Table I, particularly the ``amplitude in excited cavity" and ``noise
into output port."
\label{note:model}}

\begin{table}[t]
\caption{Mapping of the parameters in the BGKT microwave-resonator speed meter 
paper to those in this paper.}
\begin{tabular}{lll}
Parameter&BGKT&Purdue\\
\hline
signal frequency & $\omega$ & $\omega$ \\
carrier frequency & $\omega_e$ & $\omega_0$ \\
optimal frequency & $\omega_0$ & $\omega_{\rm opt}$ \\
mass of test body & $m$ & $m$ \\
characteristic length & $d$ & $L$ \\
sloshing frequency & $\Omega$ & $\Omega = c \sqrt{T_{\rm s}}/L$  \\
test-mass displacement & $\tilde{x}(\omega)$ & $\tilde{x}(\omega)$  \\
signal extraction rate\tablenote{$\tau_e^\ast$ is the relaxation time 
	of the excited resonator due to energy flowing out.} 
	 & $\delta_e = 1/2\tau_e^\ast$
	 & $\delta = cT_{\rm o}/4L$ \\
impedance of resonators\tablenote{In BGKT, both resonators have the same
	characteristic impedance, but in this interferometer, they are different.
	Consequently, caution must be used when transforming between the two models.} 
	& $\rho$ & $\rho_{\rm o} = 2L/c\sqrt{T_{\rm o}}$  \\
 & & $\rho_{\rm i} =2L/c\sqrt{T_{\rm i}}$  \\
driving amplitude\tablenote{There is a proportionality
	constant $\alpha=\sqrt{2 \hbar \sqrt{T_{\rm o}}}$ which must be
	included to get the correct dimensionality when transforming
	BGKT's equations into Purdue's notation.  For example,
	$U_0 \longleftrightarrow \alpha I_1$.} & $U_0$ & $\alpha I_1$ \\
amp. in excited cavity\tablenotemark[3] & $-q_0 = U_0/\Omega \rho$ &
	$\alpha B_1=\alpha I_1/\Omega \rho_{\rm i}$  \\
noise into output port\tablenotemark[3]$^,$\tablenote{Notice that
	the quadratures are reversed.  This is due to a difference in
	the way the models were defined.} & $\{U_{es},\ U_{ec} \}$  &
	$-\alpha\{ \tilde{\ell}_1,\ \tilde{\ell}_2 \}$  \\
sideband components\tablenotemark[3]$^,$\tablenote{Notice that
	in Purdue's notation the letter indicates the cavity and the
	numerical subscript indicates the quadrature, whereas in BGKT, the
	letter indicates the quadrature and the number indicates
	the resonator.} & $\{a_1, b_1, a_2, b_2\}$  
	& $\alpha \{ \tilde{b}_1, \tilde{b}_2, \tilde{f}_1, \tilde{f}_2 \}$ \\
output amplitude\tablenotemark[3] 
	& $\tilde{U}(\omega)$ & $\alpha \tilde{k}_\Phi (\omega)$ \\
\hline
\end{tabular}
\label{mapping}
\end{table}

We assume that ordinary vacuum enters the output port of the interferometer;
i.e.\ $\tilde{\ell}_1$ and $\tilde{\ell}_2$ are quadrature amplitudes for
ordinary vacuum.  This means [Eq. (26) of KLMTV] that their spectral densities
are unity and their cross-correlations are zero.  By noting that the homodyne
output (\ref{kPhiOut}) is proportional to
\begin{equation}
\frac{2}{L} (\tilde{x}+\tilde{x}_m )
= \tilde{h}+\frac{2}{L} (\tilde{x}_{BA}+\tilde{x}_m)
\end{equation}
and examining the dependence of $\tilde{x}_{BA}$ and $\tilde{x}_m$ on the input
vacuum $\tilde{\ell}_1$ and $\tilde{\ell}_2$, we deduce the (single-sided)
spectral density of the gravitational wave output noise $\tilde{h}$:
\begin{equation}
S_{h_n} = (h_{SQL}^{\rm speed})^2 \xi^2\;,
\label{spec}
\end{equation}
where
\begin{equation}
h_{SQL}^{\rm speed} = \sqrt{\frac{16 \hbar}{m \omega^2 L^2}} \;,
\label{sqlspeed}
\end{equation}
is the standard quantum limit (SQL) for our speed-meter interferometer,
\begin{equation}
\xi^2 = {|{\cal L}(\omega)|^2 \over 2 \Lambda^4
\sin^2 \Phi} - \cot \Phi + {\Lambda^4 \over 2 |{\cal L}(\omega)|^2} \;,
\label{xisq}
\end{equation}
is the fractional amount by which the SQL is beaten
(in units of squared amplitude), and
\begin{equation}
\Lambda^4 = {\hbar T_{\rm o} T_{\rm i} (\omega_0 I_1)^2 \over
	2 L^2 m T_{\rm s}} \;.
\end{equation}
Note that the quantity $\xi^2$ is the same (modulo
a minus sign in the definition of $\Phi$) as the
quantity $\xi^2_{\rm WB}$ in BGKT [Eq.\ (40)].

[We comment, in passing, on the SQLs that appear in the various papers:
BGKT use double-sided spectral densities and measure the velocity of a
single test body with mass $\mu$.  The corresponding standard quantum
limit for position is
\begin{equation}
(S_{x,SQL}^{\rm one\ body})_{\rm double \mbox{-}sided} = \frac{\hbar}{\mu \omega^2}
\end{equation}
[their Eq.\ (5) divided by $\mu^2 \omega^4$ to convert from force
to position and with $\mu$ denoted by $m$].
KLMTV and the present paper
used single-sided spectral densities, i.e.\ we fold negative frequencies
into positive, so our one-body SQL is
\begin{equation}
(S_{x,SQL}^{\rm one\ body})_{\rm single \mbox{-}sided}
= \frac{2 \hbar}{\mu \omega^2} \;.
\end{equation}
For our speed meter, the quantity measured is the relative velocity of
two mirrors, $x=x_1-x_2$, for which the gravitational-wave signal is
$\frac{1}{2} \tilde{h} L$ and the reduced mass is $\mu=m/2$, so our
gravity-wave SQL spectral density is
\begin{eqnarray}
(S_{h,SQL}^{\rm speed\ meter}&&)_{\rm single \mbox{-}sided}
	\equiv (h_{SQL}^{\rm speed})^2 \nonumber \\
	&& =\biggl( \frac{2}{L} \biggr)^2 \frac{2 \hbar}{(m/2)\omega^2}
	=\frac{16 \hbar}{m \omega^2 L^2} \;.
\end{eqnarray}
For a conventional interferometer, as analyzed by KLMTV, the quantity
measured is the relative position of four mirrors, $x=(x_1-x_2)-(x_3-x_4)$,
for which the gravitational-wave signal is $2 \cdot \frac{1}{2} \tilde{h} L=
\tilde{h}L$ and the reduced mass is $\mu=m/4$, so the gravity-wave
SQL spectral density is
\begin{eqnarray}
(S_{h,SQL}^{\rm conv}&&)_{\rm single \mbox{-}sided}
	\equiv (h_{SQL}^{\rm conv})^2 \nonumber \\
	&&=\biggl( \frac{1}{L} \biggr)^2 \frac{2 \hbar}{(m/4)\omega^2}
	=\frac{8 \hbar}{m \omega^2 L^2} \;,
\end{eqnarray}
half as large as for our speed meter.  If we were to build a speed meter
consisting of two excited cavities (one in each arm) and two unexcited
cavities (as in Fig.\ 4 of BGKT), then our speed meter SQL would be reduced
by a factor of 2, to the same value as for a conventional interferometer.]

Continuing with our analysis, we can express $|{\cal L}(\omega)|^2$
[Eq.\ (\ref{calL})] as
\begin{equation}
|{\cal L}(\omega)|^2 = (\omega^2-\omega_{\rm opt}^2)^2
	+ \delta^2 (\omega^2_{\rm opt} +\delta^2 /4) \;,
\end{equation}
where
\begin{equation}
\omega_{\rm opt} = \sqrt{\Omega^2 - \delta^2/2} \;,
\label{omegaopt}
\end{equation}
as we shall see, is the interferometer's optimal frequency.
These two expressions are identical to Eqs. (37) and (38) of BGKT.
We shall optimize the homodyne angle $\Phi$ to minimize the noise
at some specific frequency, $\omega_F$.  The result is
\begin{equation}
\cot \Phi = {\Lambda^4 \over |{\cal L}(\omega_F)|^2} \;.
\label{phase}
\end{equation}
Then, Eqs.\ (42)--(48) of BGKT apply exactly to the analysis here:
$\xi^2(\omega)$ for this homodyne phase $\Phi$ (\ref{phase}) is
\begin{equation}
  \xi^2 (\omega) = \frac{ |{\cal L}(\omega)|^2 }{ 2\Lambda^4 } +
    \frac{\Lambda^4(\omega^2-\omega_F^2)^2(\omega^2+\omega_F^2-2\omega_{\rm opt}^2)^2}
      {2|{\cal L}(\omega)|^2|{\cal L}(\omega_F)|^4} \;,
\end{equation}
and its minimum is
\begin{equation}
  \xi^2_{\rm min} = \xi^2(\omega_F) =
    \frac{(\omega_F^2-\omega_{\rm opt}^2)^2+\delta^2(\omega_{\rm opt}^2+\delta^2/4)}
      {2\Lambda^4}\;.
\end{equation}
The noise can be further minimized by setting the speed meter's
optimal frequency to $\omega_{\rm opt} = \omega_F$ to get
\begin{equation}
  \xi^2(\omega) = \frac{|{\cal L}(\omega)|^2}{2\Lambda^4} +
    \frac{\Lambda^4(\omega^2-\omega_{\rm opt}^2)^4}
      {2|{\cal L}(\omega)|^2\delta^4(\omega_{\rm opt}^2+\delta^2/4)^2}\; ,
  \label{xi_omega}
\end{equation}
with
\begin{equation}
  \xi^2_{\rm min}  = \frac{\delta^2(\omega_{\rm opt}^2+\delta^2/4)}{2\Lambda^4}
   = {W^{\rm SQL}_{\rm circ}\over W_{\rm circ}}\;.
  \label{xi_min}
\end{equation}
Here $W_{\rm circ}$ is the power circulating in the excited arm
\footnote{Note that that Eq.\ (\ref{xi_min}) uses the power circulating in
the excited cavity,
$W_{\rm circ}$, whereas BGKT's quantity $W$ in their Eq.\ (45) is equivalent
to the power {\it transmitted} through the interferometer's input mirror.  This
quantity $W$ is also the amount of power extracted with the signal at the
output port ($W_{\rm exit}$ in Sec.\ \ref{sec:losslessnum} and 
\ref{sec:losslessdisc}).}
[Eq. (\ref{eq:poweramp})] and
\begin{eqnarray}
W^{\rm SQL}_{\rm circ}
	&=& {mL^2 \delta^2 (\omega_{\rm opt}^2 + \delta^2/4) \over 8 \omega_0 T_{\rm o}}
		\nonumber \\
	&=& (0.8 {\rm MW}) \biggl(\frac{m}{40 {\rm kg}} \biggr)
		\biggl(\frac{L}{4 {\rm km}} \biggr)^2 \biggl(\frac{0.07}{T_{\rm o}} \biggr)
		\nonumber \\
	&& \quad \times 
		\biggl(\frac{\omega_{\rm opt}}{2\pi \times 100 {\rm Hz}} \biggr)^4
		\biggl(\frac{1.78\times 10^{15} {\rm Hz}}{\omega_0} \biggr)
\label{WSQL}
\end{eqnarray}
is the circulating power required to reach the standard quantum limit at the
optimal frequency $\omega_{\rm opt}$ (we have assumed $\delta=2\omega_{\rm opt}$
to get the second line of the above equation; see Sec.\ \ref{sec:losslessnum}).
By pumping with a power $W_{\rm circ}>W_{\rm circ}^{\rm SQL}$, the
speed meter can beat the SQL in the vicinity of the optimal frequency
$\omega_{\rm opt}$ by the amount $\xi_{\rm min}^2=W^{\rm SQL}_{\rm circ}/ W_{\rm circ}$.

If [following BGKT Eqs.\ (47) and (48)] we define the frequency
band $\omega_1 < \omega < \omega_2$ of high
sensitivity to be those frequencies for which
\begin{equation}
  \xi(\omega) \le \sqrt2\xi(\omega_{\rm opt})\;, \label{range}
\end{equation}
then Eqs.\ (\ref{xi_omega}) and ({\ref{xi_min}) imply that
\begin{eqnarray}
  \omega_{1,2}^2 &=& \omega_{\rm opt}^2 \mp
    \frac{\delta^2(\omega_{\rm opt}^2+\delta^2/4)}
      {\sqrt[4]{\delta^4(\omega_{\rm opt}^2+\delta^2/4)^2+\Lambda^8}} \nonumber \\
      &=& \omega_{\rm opt}^2 \mp
      	\frac{2\Lambda^2\xi_{\rm min}^2}{\sqrt[4]{4\xi_{\rm min}^4+1}} \;.
\label{omega12}
\end{eqnarray}
Equations (\ref{omega12}), (\ref{xi_omega}), and
(\ref{xi_min}) imply that the lossless
speed meter can
beat the force-measurement SQL by a large amount $\xi_{\rm min}\ll 1$ over a wide
frequency band, $\omega_1 \ll \omega_2 \sim \sqrt{2} \omega_{\rm opt}$ by setting
\begin{equation}
\frac{\Lambda}{\omega_{\rm opt}} \sim \frac{\delta^2}{2 \omega_{\rm opt}^2}
	\agt 2 \;.
\label{WBreq}
\end{equation}
A plot of $\xi^2$, optimized in this manner but for rather modest parameter
values, is shown in Fig.\ \ref{fig:xisq}.

\begin{figure}
\epsfig{file=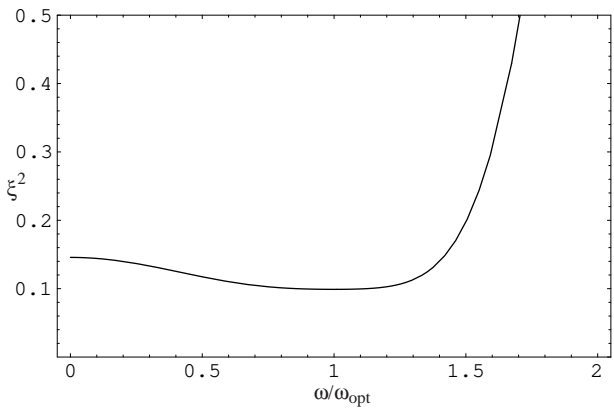}
\caption{Plot of the squared amount by which the speed meter beats the standard 
quantum limit ($h_{SQL}^{\rm speed}$), as a function of frequency 
(normalized to the optimal frequency, $\omega_{\rm opt}$).  For the 
parameter values $\xi_{\rm min}^2=0.1$,
$\delta=2\omega_{\rm opt}$, and $\Lambda^4=40 \omega_{\rm opt}^4$, this 
is identical to the speed meter curve in Fig.\ 3 of BGKT.}
\label{fig:xisq}
\end{figure}

\subsection{Numerical Analysis}
\label{sec:losslessnum}

To get an idea of the magnitudes of the quantities involved in this
interferometer, we can start by
combining the wide-bandwith requirement (\ref{WBreq}) with the
definitions $\delta = c T_{\rm o}/4L$,
$\Omega = c \sqrt{T_{\rm s}}/L$, and $\omega_{\rm opt}^2 = \Omega^2 -\delta^2/2$.  From these, we find that the wide-bandwidth
requirement $\delta^2 \agt 4 \omega_{\rm opt}^2$ becomes
$T_{\rm o}^2 \agt (64/3) T_{\rm s}$.  If, as in BGKT,
we take $\delta=2\omega_{\rm opt} =2\omega_F$ but set
$\omega_{\rm opt}=2\pi \times 100 {\rm Hz}$ as in KLMTV,
then that gives $T_{\rm o}=0.07$ and $T_{\rm s}=0.0002$.  Notice
this particular value of $T_{\rm s}$ does not satisfy the condition
$\omega_{\rm opt} \ll \Omega$, which was necessary
to get a signal that is {\it only}
proportional to the velocity of the test masses' motion.  Instead,
we have $\omega_{\rm opt} \sim \Omega$, which 
implies that the signal consists
of a linear combination of odd time derivatives of position, with
substantial contributions coming from derivatives higher than the speed
[see Eq. (\ref{kPhiOut})].

If, in addition to $\delta=2\omega_{\rm opt}=2\omega_F=4 \pi \times 100 {\rm Hz}$,
we choose $\xi_{\rm min}^2 = 0.1$ (as in BGKT), then we find
$W_{\rm circ}^{SQL} \simeq 0.8\ {\rm MW}$ from Eq.\ (\ref{WSQL}) and
a circulating power of $W_{\rm circ}\simeq 8\ {\rm MW}$.
The input-port transmissivity
is not explicitly defined by the above
requirements, but it is required, in our analysis, to be much
smaller than $T_{\rm s} = 0.0002$ or $T_{\rm o}=0.07$, i.e.
$T_{\rm i} \lesssim 2 \times 10^{-5}$.  This
then dictates an outrageously high input power of $\agt$300 MW to get the
needed circulating power.  The power that exits through the signal port,
along with the signal, is
$W_{\rm exit}= T_{\rm o} W_{\rm circ}\sim 0.56\ {\rm MW}$.
The resulting noise curve is shown in Fig.\ \ref{fig:noisecurve}; the
parameter values used are given in Table \ref{table:params}.

\begin{figure}
\epsfig{file=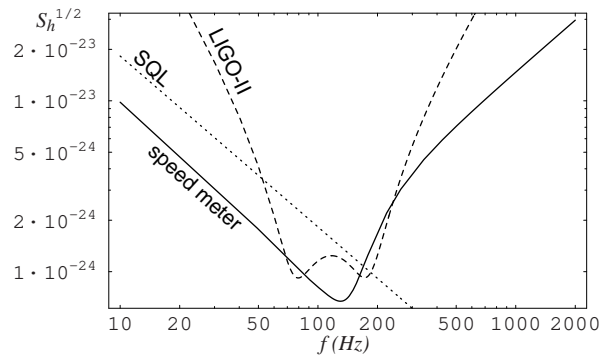}
\caption{Lossless noise curve for a speed meter 
optimized at a 
frequency of 100 Hz.  The transmissivities and power are given in Table
\ref{table:params}. The dashed line represents the theoretical LIGO-II
noise curve in which a signal-recycling mirror and optical noise correlations
have been used to beat the SQL (and thermal noise has been made negligible),
 as described by Buonanno and Chen 
\protect\cite{SRrefa,SRrefb}.  The dotted line represents the SQL; we use
$h_{SQL}^{\rm conv}$ because we are comparing to a position meter.}
\label{fig:noisecurve}
\end{figure}

\begin{table}[t]
\caption{Interferometer parameters and their fiducial values.}
\begin{tabular}{lll}
Parameter&Symbol&Fiducial Value\\
\hline
carrier frequency & $\omega_0$ & $1.78 \times 10^{15} {\rm s}^{-1}$ \\
mirror mass & $m$ & 40 kg \\
arm length & $L$ & 4 km \\
sloshing mirror transmissivity & $T_{\rm s}$ & 0.0002 \\
input mirror transmissivity & $T_{\rm i}$ & $2 \times 10^{-5}$ \\
output mirror transmissivity & $T_{\rm o}$ & 0.07 \\
end mirror transmissivity & $T_{\rm e}$ & $2 \times 10^{-6}$ \\
SQL circulating power & $W_{\rm circ}^{SQL}$ & 1.7 MW \\
\hline
\end{tabular}
\label{table:params}
\end{table}

Recall that this analysis is for only
one speed meter, which is equivalent to a single arm of the conventional
LIGO design.  If we were to add another speed meter (another pair of
cavities) with the position of the excited and unexcited cavities reversed,
interfering the output beams would increase the
sensitivity by a factor of two,
in much the same way as having two arms increases the sensitivity
in the conventional LIGO design.  In addition, doing this would reduce
the interferometer's sensitivity to laser frequency fluctuations in the same
way as having two arms in conventional LIGO designs.

\subsection{Discussion of Lossless Speed Meter}
\label{sec:losslessdisc}

In this section, we will look at variety of issues that should be
understood and addressed in the process of developing a
different, more practical, speed-meter design.  One problem is the
large circulating power ($\sim$8 MW) required to achieve wide-band
sensitivity a factor $\sim 10$ in noise power below the standard quantum limit.
A second problem is how to get that light into the interferometer,
as the present design requires an input power that is outrageously high.
This is, at least partly, the result of the high reflectivity
of the input mirror, which causes most of the input light to be
reflected back towards the laser.

A third problem is the amount of
power flowing through the interferometer:  With a circulating power of
$\sim$8 MW, the power extracted with the signal is
$W_{\rm exit} = T_{\rm o} W_{\rm circ} \sim 0.56\ {\rm MW}$. This
same amount of power must be fed into the excited cavity to maintain
a steady state.  To reduce this power through-put, we could
decrease $T_{\rm o}$ substantially; however, doing this will cause the
wide-bandwidth requirement (\ref{WBreq}) to be violated, and consequently,
the behavior of the speed meter will become more narrow-band.
In fact, the effect of changing $T_{\rm o}$ is strong enough that if it is
decreased by one order of magnitude, the speed meter will
no longer beat the SQL except for a very narrow range of frequencies.
This clearly is not a viable solution.

Another approach, in which this large through-put power
might conceivably be tolerated, is to recycle it back into the interferometer.
To do that, we must strip the signal off by using a beamsplitter to
interfere the outputs from two speed-meter
interferometers as in Fig.\ 4 of BGKT.  Using this ``double" speed meter
could also help increase the sensitivity, as described at the end of the previous
section.

Turning to the issue of the high circulating power, it should first be noted that
the circulating power required to reach the SQL, $W_{\rm circ}^{SQL} \sim
800\ {\rm kW}$ is comparable to that for conventional interferometers
[Eq.\ (132) of KLMTV gives $W_{\rm circ}^{SQL} \sim
840\ {\rm kW}$ with $m=40 {\rm kg}$, instead of 30kg].  
A double speed meter, as described above, would have twice the sensitivity
as a single speed meter at the same power.
As mentioned in Sec.\ \ref{sec:Introduction},
the high powers needed to reach or beat the SQL are a common feature of
many QND designs \cite{highpower}, for example the variational-output
interferometer discussed in KLMTV.

A likely method of reducing the needed circulating power, without losing
the wide-band performance of the speed meter, is to inject squeezed vacuum
into the output port, as was originally proposed by Caves \cite{Cavessqueezedvacuum}
for conventional interferometers and by KMLTV for their QND
squeezed-input and squeezed-variational interferometers.  In these cases,
for realistic amounts of squeezing, the circulating power can be
reduced by as much as an order of magnitude
\cite{Cavessqueezedvacuum,Kimble}.  Detailed analyses applying
squeezed-vacuum techniques to speed meters have
not yet been carried out, but if the effect is similar, it would have the
beneficial side-effect of reducing the needed input power by the same
amount, which might be useful in a redesigned speed meter.

As for the outrageously high input power, the fact that so much of the light
impinging on the input-port mirror is
reflected back to the laser suggests an obvious solution would be adding a
power-recycling mirror and/or increasing the transmissivity $T_{\rm i}$
of the input mirror.  However,
neither of these approaches addresses the fundamental problem: there is
an empty cavity between the driving laser and the excited
cavity.  In a conventional LIGO-type interferometer, the laser
drives a strongly excited Fabry-Perot cavity directly. In that case, Bose
statistics dictate that photons will be ``sucked" into the cavities,
producing a strong amplification.  Hence, there will be significantly
more power stored in the arms of the interferometer than the
driving laser is producing.  Without losses,
\begin{equation}
\frac{\rm circulating\ power}{\rm input\ power} \sim
\frac{8}{T_{\rm PR} T_{\rm IM}} \sim 10^5\;,
\end{equation}
where $T_{\rm PR} \sim 0.06$ is the transmissivity of the
power-recycling mirror and
$T_{\rm IM}\sim 0.005$ is that of the internal mirrors \cite{LIGOIIdoc}.
However, in this speed meter
design, there is an empty cavity instead of a
low power laser feeding into the highly-excited cavity so Bose
statistics do not help us.  The
result is the need for a driving laser that produces far
more power than is stored in the arms of the speed meter:
\begin{equation}
\frac{\rm circulating\ power}{\rm input\ power} \sim
\frac{T_{\rm i}}{4 T_{\rm s}} \sim 10^{-3}\;.
\end{equation}

One way to address this problem would be to allow a small amount of
light to build up in the previously empty cavity.  This would
cause position information to contaminate the previously pure
speed meter behavior.  However, this solution is not ideal because,
as the amount of light in the ``empty"
cavity increases, the sensitivity degrades faster than the
required input power decreases.  To consider this more closely,
we first need to remove the restriction (\ref{eq:darkdriving})
on the light $L_1$ fed into the
output port, which forces the unexcited cavity
[denoted by $F(\zeta)$ and $G(\eta)$ in Fig.\ 1]
to be truly empty.  Instead, we let $L_1$ be determined by the amount
of power we want to have in the unexcited cavity.
Secondly, since the unexcited cavity is no longer empty, we need to include the
movement of the end mirror in that cavity, which we previously neglected.
This requires revising Eqs.\ (\ref{eqs:Geq}) to include $\tilde{x}$ terms
(with back action) as in Eqs.\ (\ref{eqs:Ceq}).  To calculate
how much the needed input power decreases as a function of
the ratio of the powers of the two cavities, we can solve for the carrier,
as in Eqs. (\ref{allcarriers}), and do some algebra to express
the input amplitude $I_1$ as a function of the excited-cavity amplitude
$B_1$ and the ratio of the amplitudes of the powers of the two cavities
($F_1/B_1$).  After converting from amplitudes to powers, the relationship
between the input powers is
\begin{equation}
{W_{\rm input} (R) \over W_{\rm input} (R=0)}
	= \Biggl[ 1-\frac{T_{\rm i}\sqrt{R}}{2 \sqrt{T_{\rm s}}} \Biggr] \;,
\end{equation}
where $R$ is the ratio of the powers in the two cavities.
Since we require $T_{\rm i} \ll T_{\rm s} \ll 1$ and $R \ll 1$ to get
speed-meter behavior, $W_{\rm input}$ cannot be reduced much at all.

Also of concern here
is how much position information will be included in the output.
To calculate the strength of the position signal, relative to the strength of
the velocity signal, we can solve the revised equations (as described in the
previous paragraph) for the sideband-light output.  Then taking the ratio of
the coefficients of the position $\tilde{x}$ term
and the velocity $\tilde{x}$ term, we find
\begin{equation}
\Bigl| \frac{\rm position}{\rm velocity} \Bigr| \sim
\frac{c \sqrt{T_{\rm s}}}{\omega L} \sqrt{R}
=\frac{\Omega}{\omega} \sqrt{R} \sim \sqrt{R} \;.
\label{eq:poscont}
\end{equation}
Since the spectral density involves the square of the amplitudes used to
calculate the above expression (\ref{eq:poscont}), $S_{h_n}$ and $\xi^2$
will scale with $R$.  This indicates that even a modest amount of
power in the `empty' cavity will introduce
a significant amount of position information into the output signal.
The effect of this, for a few values of $R$, can be seen in
Fig.\ \ref{fig:poscont}.

\begin{figure}
\epsfig{file=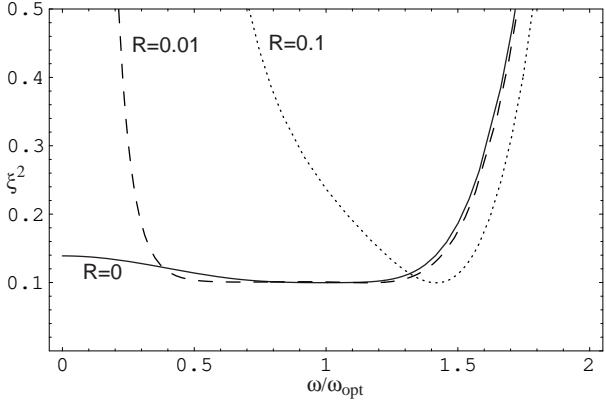}
\caption{Plot of $\xi^2$ for the lossless speed meter optimized
at a frequency of 100 Hz, with varying amounts of power in the `empty' cavity.
The $R=0$ curve is the same as that in Fig. \ref{fig:xisq}.  
Transmissivities are $T_{\rm s}=0.0002$, $T_{\rm o}=0.07$,
$T_{\rm i}=2\times 10^{-5}$, $T_{\rm e}=0$, and the circulating power
is 17 MW.}
\label{fig:poscont}
\end{figure}

In fact, it appears that this problem of outrageously high input power
is the fatal flaw of this particular speed-meter design.
Yanbei Chen \cite{altspeedmeter}
has conceived a class of alternative speed meter designs that may solve this
problem.  Chen and the author are carrying out an analysis and
optimization of them; we shall report the details in a future paper.

\section{Sensitivity of Speed Meter with Losses}
\label{sec:lossy}

In order to understand the issue of optical losses and dissipation in
this type of interferometer, we shall return to the full equations presented
in Sec.\ \ref{sec:math}.  In that case, the output of the system is:
\begin{mathletters}
\begin{eqnarray}
\tilde k_1 &=& - {{\cal L}^*(\omega)\over{\cal L}(\omega)}\tilde \ell_1
+ {c^2 \sqrt{T_{\rm s} T_{\rm o}T_{\rm i}} \over 2 L^2 {\cal L}(\omega)}
	\tilde{i}_1 \nonumber \\
&& \quad - {i \omega c \sqrt{T_{\rm o} T_{\rm e}} \over 2 L {\cal L}(\omega)}
	\widetilde{m}_1
+ {c^2  \sqrt{T_{\rm s} T_{\rm o} T_{\rm e}}\over2 L^2 {\cal L}
	(\omega)} \tilde{n}_1 \;,
\label{k1outlossy} \\
\tilde{k}_2 &=&
   \frac{-i \omega \omega_0 I_1 \sqrt{T_{\rm o} T_{\rm i}}}
   	{2L \sqrt{T_{\rm s}} {\cal L}(\omega)}\tilde{x}
-\frac{{\cal L}^\ast (\omega)}{{\cal L}(\omega)}  \tilde{\ell}_2
+\frac{c^2 \sqrt{T_{\rm s} T_{\rm o}T_{\rm i}}}{2L^2 {\cal L}(\omega)}
	\tilde{i}_2 \nonumber \\
&& \quad -\frac{i\omega c \sqrt{T_{\rm o}T_{\rm e}} }{2L
	{\cal L}(\omega)}\widetilde{m}_2
+\frac{c^2 \sqrt{T_{\rm s} T_{\rm o}T_{\rm e}}}{2L^2 {\cal L}(\omega)}
	\tilde{n}_2\;,
\label{k2outlossy}
\end{eqnarray}
\end{mathletters}
where
\begin{eqnarray}
\tilde{x} =  \frac{1}{2} L\tilde{h}
   &+& \frac{\hbar \omega_0 I_1}{m \omega^2 L^2 {\cal L}(\omega)} \Biggl[
   	\frac{i \omega L \sqrt{T_{\rm o} T_{\rm i}}}{\sqrt{T_{\rm s}}}
   	  \tilde{\ell}_1
   - c T_{\rm i} \tilde{i}_1 \nonumber \\
   && \quad \quad + \frac{ i\omega L \sqrt{T_{\rm i} T_{\rm e}}}
   { \sqrt{T_{\rm s}} } \widetilde{m}_1
   - c \sqrt{T_{\rm i} T_{\rm e}} \tilde{n}_1  \Biggr]\;.
\end{eqnarray}
As before, we can express these in a more concise way:
\begin{mathletters}
\begin{eqnarray}
\tilde{k}_1 &=& \tilde{\ell}_1 e^{2 i \psi}
	-(\tilde{i}_1 \kappa_i+\tilde{n}_1 \kappa_n) e^{i \theta}
		+\widetilde{m}_1 \kappa_m e^{i \psi} \;, \\
\tilde{k}_2 &=& \bigl(\tilde{\ell}_2-\kappa \tilde{\ell}_1 
		- \kappa \frac{\sqrt{T_{\rm e}}}{\sqrt{T_{\rm o}}} 
			\widetilde{m}_1 \bigr) e^{2i \psi} 
 -(\tilde{i}_2 \kappa_i +\tilde{n}_2  \kappa_n) e^{i \theta} \nonumber \\
	&&\quad + \biggl(\frac{\sqrt{\kappa}}{h_{SQL}^{\rm conv}} \tilde{h}
	 +\kappa_m \widetilde{m}_2 \biggr) e^{i \psi} 
	 \nonumber \\
	 && \quad +\biggl( \tilde{i}_1 \sqrt{T_{\rm i}}
			+ \frac{\sqrt{T_{\rm e}}}{\sqrt{T_{\rm o}}} \tilde{n}_1 \biggr)
			\frac{\Omega}{\omega} \kappa e^{i \phi} \;,
\end{eqnarray}
\label{ksoutlossyKimbleform}
\end{mathletters}
where, in addition to the definitions given by Eqs.\ (\ref{Kimbleparams}),
\begin{equation}
\tan \theta = - \cot \psi \;, \quad \tan \phi = -\cot 2\psi \;,
\end{equation}
and
\begin{mathletters}
\begin{eqnarray}
\kappa_i &=& \sqrt{\frac{c^4 T_{\rm s} T_{\rm o} T_{\rm i}}
	{4 L^4 |{\cal L}(\omega)|^2}}\;, \\
\kappa_m &=& \sqrt{\frac{c^2 T_{\rm o} T_{\rm e} \omega^2}
	{4 L^2 |{\cal L}(\omega)|^2}}\;, \\
\kappa_n &=& \sqrt{\frac{c^4 T_{\rm s} T_{\rm o} T_{\rm e}}
	{4 L^4 |{\cal L}(\omega)|^2}} \;.
\end{eqnarray}
\end{mathletters}

Once again, we do homodyne detection and calculate the spectral
density of the noise.  (It should be noted that, in the lossy case,
 there are enough
differences between the optical speed meter and the BGKT microwave
speed meter to obscure the mapping.  Consequently, we will not be able
to present as close a comparison in this section as we did in the lossless case.)
The fractional amount by which the SQL is beaten is
\begin{equation}
\xi^2 =
	{|{\cal L}'(\omega)|^2 \over 2 \Lambda^4 \sin^2 \Phi}
	- \cot \Phi
	+{\Lambda^4 \over 2 |{\cal L}(\omega)|^2} \;,
\label{lossyspec}
\end{equation}
where
\begin{equation}
|{\cal L}'(\omega)|^2 = (\omega^2 - \omega'^2_{\rm opt})^2
	+ \delta \delta^*[\omega'^2_{\rm opt} + \frac{\delta \delta'}{4 \delta^*}
	(\delta^* +\delta_e+\delta_i)] \;,
\end{equation}
with
\begin{eqnarray}
\delta_i &=& c T_{\rm i}/L \;, \\
\delta_e &=& c T_{\rm e}/L \;, \\
\delta' &=& \delta + \delta_e \;, \\
\delta^* &=& \delta + 2 \delta_e +\delta_i \;,
\end{eqnarray}
and
\begin{equation}
\omega'_{\rm opt} = \sqrt{\Omega^2 - \delta \delta'/2} \;.
\end{equation}
Optimizing the homodyne angle at frequency $\omega_F$ gives
\begin{equation}
\cot \Phi = {\Lambda^4 \over |{\cal L}'(\omega_F)|^2} \;.
\label{lossyphase}
\end{equation}
The resulting $\xi^2$ is
\begin{equation}
\xi^2 =
	{|{\cal L}'(\omega)|^2 \over 2 \Lambda^4}
	- {\Lambda^4 \over |{\cal L}'(\omega_F)|^2}
	+ {\Lambda^4 |{\cal L}'(\omega)|^2 \over 2 |{\cal L}'(\omega_F)|^2}
	+{\Lambda^4 \over 2 |{\cal L}(\omega)|^2} \;.
\end{equation}
Setting $\omega=\omega_{\rm opt}'=\omega_F$ gives
\begin{eqnarray}
\xi^2 (\omega_{\rm opt}') =&&
	{\delta^2 \delta' (\delta_e +\delta_i )/4 + \delta \delta^* (\omega_{\rm opt}'^2
		+\delta \delta'/4) \over 2 \Lambda^4} \nonumber \\
	&& -{\Lambda^4 \over 2 [\delta^2 \delta' (\delta_e +\delta_i )/4 + \delta \delta^*
		 (\omega_{\rm opt}'^2 +\delta \delta'/4)]} \nonumber \\
	&& +{\Lambda^4 \over 2 [(\omega_{\rm opt}'^2 - \omega_{\rm opt}^2)^2 +
		\delta^2 (\omega_{\rm opt}^2 + \delta^2 /4)]} \;.
\end{eqnarray}
Cf. Eq.\ (\ref{xi_min}).  Note that, as in BGKT, the sensitivity 
in the lossy case does not continue to
grow indefinitely with the growth of the parameter $\Lambda$.

Despite the presence of the additional terms included to account for
losses, the speed meter curve is largely unchanged if
we maintain our assumptions about the relative sizes of the transmissivities (\ref{transreq}).  In fact, the only losses that contribute significantly
are those associated with 
$\tilde{i}_1$ (i.e., noise entering the bright port along with the laser
light).   This term causes the speed meter to become less
sensitive at frequencies $\ll \omega_{\rm opt}$, as seen in
Fig.\ \ref{fig:lossyxisq}.  Since that is roughly the
frequency at which seismic noise becomes dominant, the effect of more
limited sensitivity in that range is not important.

\begin{figure}
\epsfig{file=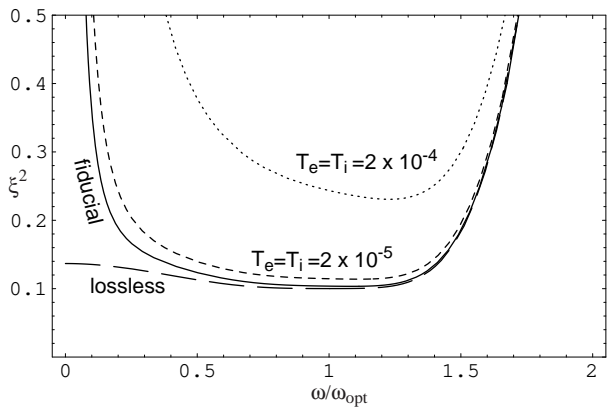}
\caption{Plot of $\xi^2$ for the speed meter with losses.  The solid curve
uses the transmissivities given in Table \ref{table:params}.  The lossless
curve has $T_{\rm e}=0$, as in Fig.\ \ref{fig:xisq}.  The other two curves
differ from the fiducial-value curve only by the specified transmissivities.}
\label{fig:lossyxisq}
\end{figure}

As it turns out, the equations in this section are valid into the regime where
$T_{\rm e} \simeq T_{\rm i} \simeq T_{\rm s}$.  In that case, 
the $\tilde{n}_1$ term will be the same size as the $\tilde{i}_1$ term, and
together, they become dominant at frequencies $\lesssim \omega_{\rm opt}$, 
while the rest of the loss terms continue to be
insignificant for this parameter regime.
Presumably, the sensitivity degradation by $\tilde{n}_1$ and $\tilde{i}_1$
are the result of vacuum fluctuations entering
into the empty cavity and contaminating the `sloshing' light.  
This behavior is shown in Fig.\ \ref{fig:lossyxisq}.
As can be seen from that plot, the interferometer loses
wideband sensitivity when operating in this regime.

\section{Conclusions}

We have analyzed the speed-meter interferometer proposed by BGKT and
have shown that it does, indeed, measure test-mass speed (and time 
derivatives of speed) rather than test-mass position.  We have also 
shown that it is capable of beating the
SQL over a broad range of frequencies.  However, the very high circulating
and input powers it requires render this design
impractical for use in LIGO-III.  It is possible, however, that there
are variations of this design that will be more feasible.

There are three separate but related problems related to the laser
power involved in this speed meter.  One is the amount of circulating
power ($\sim$8 MW) required to beat the SQL substantially (by a factor
10 in noise power) over a 
wide range of frequencies.  Another is the amount of power
coming out of the interferometer with the signal ($\sim$0.56 MW).
Both of these are serious problems, but there are conceivable
solutions to them.  The third and most severe problem
is the fact that the excited cavity is being fed through an
empty cavity.  This dramatically increases the amount of input
power needed to achieve a given circulating power, to the point where
the input is significantly greater than the circulating power.

Motivated by what we have learned in this analysis, Yanbei Chen and 
the author are developing and exploring alternative designs for speed-meter
interferometers that may solve the above problems and actually be
practical.

\section*{Acknowledgments}

I thank Kip Thorne for proposing this research problem and for helpful
advice about its solution and about the prose of this paper. 
This research was supported in part by NSF grant PHY-0099568.

\end{document}